\documentclass[12pt]{iopart}
\usepackage{graphicx}
\usepackage{overpic}
\usepackage{xcolor}
\definecolor{light-gray}{gray}{0.65}
\begin{document}

\title[Search of critical effects in NA61/SHINE]{Search of critical effects in NA61/SHINE}

\author{Magdalena Kuich for the NA61/SHINE Collaboration}

\address{Faculty of Physics, University of Warsaw, Warsaw, Poland}
\ead{mkuich@fuw.edu.pl}
\vspace{10pt}
\begin{indented}
\item[]August 2019
\end{indented}

\begin{abstract}
NA61/SHINE is a multi-purpose experiment to study hadron-proton, hadron-nucleus and nucleus-nucleus collisions at the CERN Super Proton Synchrotron (SPS). The experiment performs unique measurements for the physics of strong interactions as well as important reference measurements for neutrino and cosmic-ray physics.

The primary goals of the experiment are the study of the onset of deconfinement and the search for the critical point of the strongly interacting matter, to uncover the mechanism of thermalisation and to test the validity of statistical models. For this purpose, a two-dimensional scan was performed by varying the beam momentum (13$A$--158$A$~GeV/$c$) and the size of colliding systems (p+p, p+Pb, Be+Be, Ar+Sc, Xe+La, Pb+Pb).

In this contribution, we present recent NA61/SHINE results on the search for critical effects in spectra and fluctuations. We see no clear indication for the onset of deconfinement for intermediate colliding systems. Nevertheless, results from p+p interactions on spectra and onset of deconfinement reveal anomalous behaviour in proximity to transition energy. In fluctuation analysis of collisions of medium size nuclei at the top SPS energy, no prominent signal of critical point was observed.
\end{abstract}

%
\vspace{2pc}
%
\submitto{\PS}
%
\maketitle
%
%

\section{Introduction}
NA61/SHINE  is a multi-purpose facility measuring hadron production in hadron-proton, hadron-nucleus and nucleus-nucleus collisions. It is a fixed target experiment located in the H2 beam-line of CERN's SPS North Area.

NA61/SHINE performed a two-dimensional scan in collision energy (13A-150$A$ GeV/$c$ and system size (p+p, p+Pb, Be+Be, Ar+Sc, Xe+La, Pb+Pb) to study the phase diagram of strongly interacting matter. The main goals of NA61/SHINE are the search for the critical point and a study of the onset of deconfinement.

\section{NA61/SHINE facility}
\begin{figure}[h]
\centering
\includegraphics[width=15.8cm, trim={0 0 0 1.2cm}, clip]{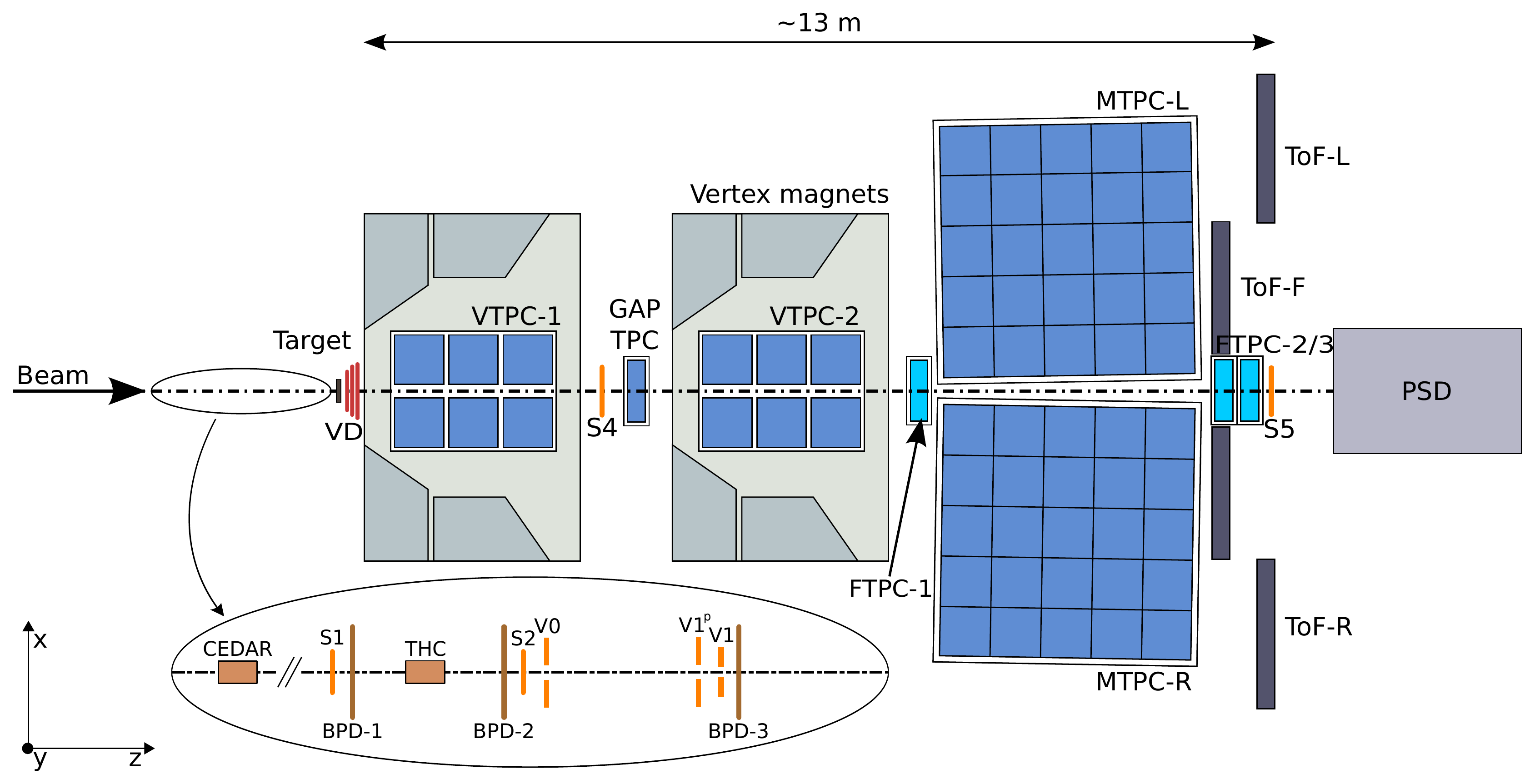}
\caption{Updated schematic layout of the NA61/SHINE detection system \cite{facilitypaper}.}
\label{fig:detector}
\end{figure}

The NA61/SHINE detection system, presented in Fig. \ref{fig:detector}, is a large acceptance hadron spectrometer with excellent capabilities in measurements of charged particles. The spectrometer is based on a set of eight Time Projection Chambers, two of which are placed in the superconducting magnets, complemented by three Time-of-Flight detectors. This setup allows for precise momentum reconstruction and identification of charged particles.
The high-resolution forward calorimeter, the Projectile Spectator Detector (PSD), measures energy flow around the beam direction, which in nucleus-nucleus reactions is primarily a measure of the number of the projectile spectators, i.e non-interacting projectile nucleons. Thus, the measurement of so-called Forward Energy (FE) can be related to the violence (centrality) of the collision. The incoming beam is monitored by a set of beam detectors, which allows to identify beam particles and measure precisely their trajectories \cite{facilitypaper}.

\section{Study of the onset of deconfinement}
The Statistical Model of the Early Stage (SMES) \cite{onset} predicts a 1$^{\textup{st}}$ order phase transition from the Quark-Gluon Plasma (QGP) to a Hadron-Resonance Gas (HRG) \cite{hrg} phase in the energy range available at the SPS.  According to SMES, at low collision energies, only pure HRG is produced, at higher – pure QGP state. The model also assumes that in some collision energy region, the transition region, both states coexist forming a mixed-phase. In the transition region, constant temperature and pressure, as well as an increase of the internal numbers of degrees of freedom, are expected. The phase transition phenomenon should manifest itself in rapid changes in the energy dependence of several hadron production properties.

\subsection{Inverse slope parameter - ``STEP''}
One of the signatures of the 1$^\textup{st}$ order phase transition, predicted by SMES due to the presence of a mixed phase of HRG and QGP, is a plateau (”step”) in the energy dependence of the so-called inverse slope parameter ($T$). SMES assumes that the energy density at the early stage of the collision increases with increasing collision energy, which leads to a rise of the initial temperature and pressure. It also results in an increase of the transverse expansion of the produced matter, visible in increasing values of the inverse slope parameter in the energy regions, in which a pure confined phase ($\sqrt{s_\textup{NN}} < 7.62$ GeV) or pure deconfined phase ($\sqrt{s_\textup{NN}} > 11.03$ GeV) is produced. In the mixed phase region (7.62 GeV $< \sqrt{s_\textup{NN}} <$ 11.03 GeV) the flattening of the inverse slope parameter energy dependence is expected due to approximately constant initial temperature and pressure. The inverse slope parameter is sensitive to both the thermal and collective motion in the transverse direction, hence an increase of the T value due to increase of kinematic freeze-out temperature and/or collective effects is expected, which is visible while increasing the size of the colliding system. \cite{onset, pedestrians}

\begin{figure}[h]
\centering
 \begin{overpic}[width=8cm]{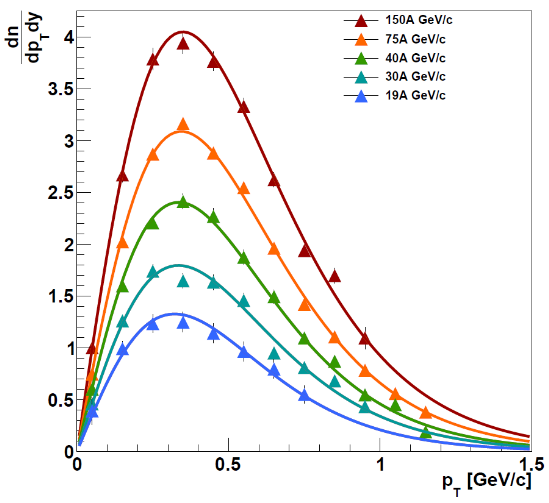}
 \put(55, 58){\scriptsize{Ar+Sc$\rightarrow$K$^+$+X}}
\end{overpic}
\caption{An example of transverse momentum spectra of K$^+$ produced in Ar+Sc collisions at five beam momenta. Solid lines illustrate spectrum parametrisation with Eq. \ref{eq:pt_spec}.}
 \label{fig:kp_spec}
\end{figure}

The $T$ parameter is obtained from parametrisation of the transverse mass ($m_\textup{T}$) / transverse momentum ($p_\textup{T}$) spectra of positively and negatively charged kaons. An example of K$^+$ $p_\textup{T}$ spectra for Ar+Sc collisions for various beam momenta is presented in Fig. \ref{fig:kp_spec}.
The kaon spectra parametrisation can be written as:
\begin{equation}
 \frac{d^2n}{dp_\textup{T} dy} = \frac{S p_\textup{T}}{T^2+T m_\textup{K}} exp \left( - \frac{\sqrt{p^2_\textup{T}+m^2_\textup{K}}}{T} \right),
 \label{eq:pt_spec}
\end{equation}
where $T$ stands for the inverse slope parameter, $m_\textup{K}$ is kaon mass and $S$ a normalisation factor. 

\begin{figure}[h]
   \begin{overpic}[width=7.9cm, trim={0.1cm 0.5cm 0.8cm 0}, clip]{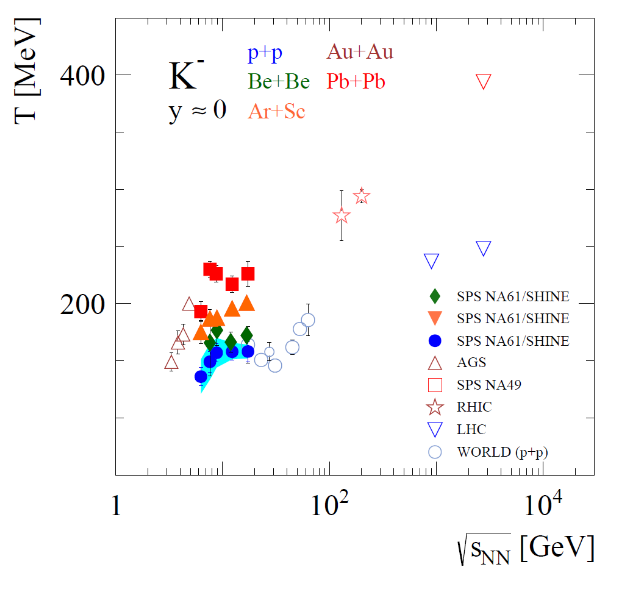}
         \put(28, 68){\scriptsize{\textbf{\textcolor{light-gray}{NA61/SHINE preliminary}}}}
    \end{overpic} 
\begin{overpic}[width=7.9cm, trim={0.1cm 0.5cm 0.8cm 0}, clip]{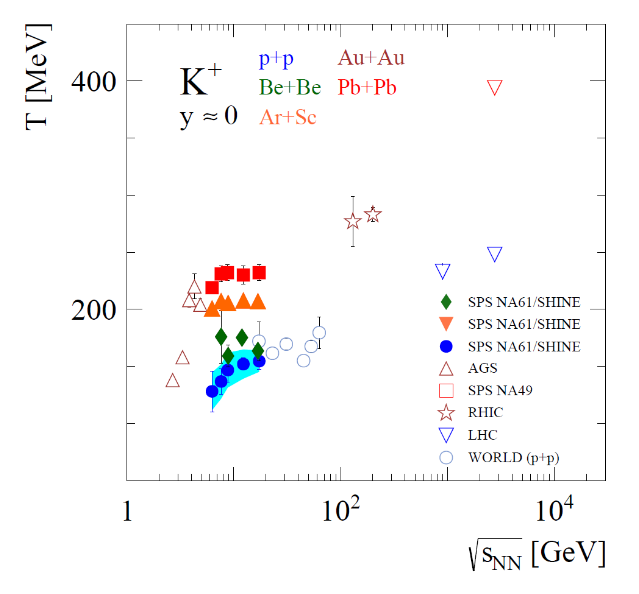}
           \put(30, 68){\scriptsize{\textbf{\textcolor{light-gray}{NA61/SHINE preliminary}}}}
    \end{overpic} 
    \caption{Inverse slope parameter $T$ of $m_\textup{T}$ spectra of K$^-$ (\textit{left}) and K$^+$ (\textit{right}) as function of collision energy. Most results are shown with statistical uncertainties only. For the p+p data the shaded band indicates systematic uncertainties.}
    \label{fig:step}
\end{figure}

The ``step'' was originally observed by the NA49 experiment in Pb+Pb collisions for $m_\textup{T}$ spectra of K$^\pm$ \cite{na49_onset}. Continuing this research, the NA61/SHINE observed qualitatively similar energy dependence in p+p interactions and such behaviour seems to emerge also in Be+Be reactions, as visible in Fig. \ref{fig:step}. The values of the $T$ parameter in Be+Be collisions are slightly above those in p+p interactions. The $T$ parameter in Ar+Sc reactions is found between those in p+p/Be+Be and Pb+Pb collisions.

\subsection{K$^+$/$\pi^+$ ratio - ``HORN''}
The other very important signature of the onset of deconfinement is the so-called ``horn''. It is expected for the energy dependence of the K$^+$/$\pi^+$ yields ratio which can be interpreted as strangeness to entropy ratio. According to SMES predictions, K$^+$/$\pi^+$ steeply rises in the HRG phase, reaches the maximum at the onset of deconfinement, decreases in the mixed-phase to an almost constant value in the QGP phase.

\begin{figure}[h]
  \begin{overpic}[width=7.9cm]{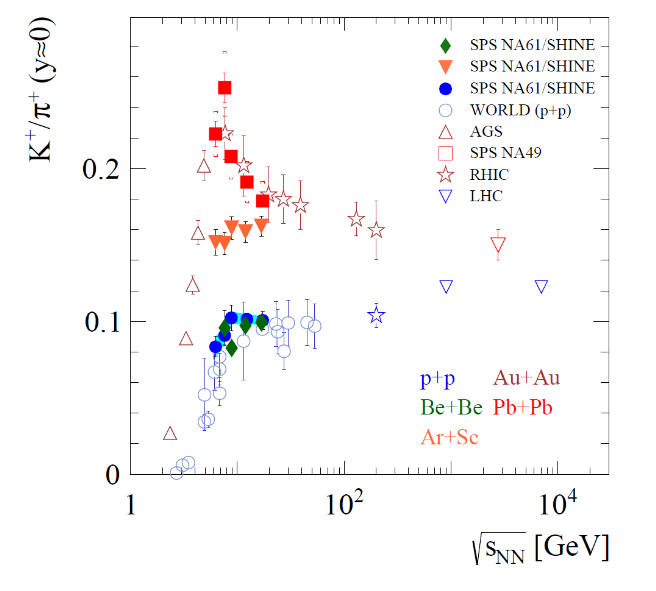}
   \put(20, 83){\scriptsize{\textbf{\textcolor{light-gray}{NA61/SHINE}}}}
      \put(20, 79){\scriptsize{\textbf{\textcolor{light-gray}{preliminary}}}}
\end{overpic}
\begin{overpic}[width=7.8cm]{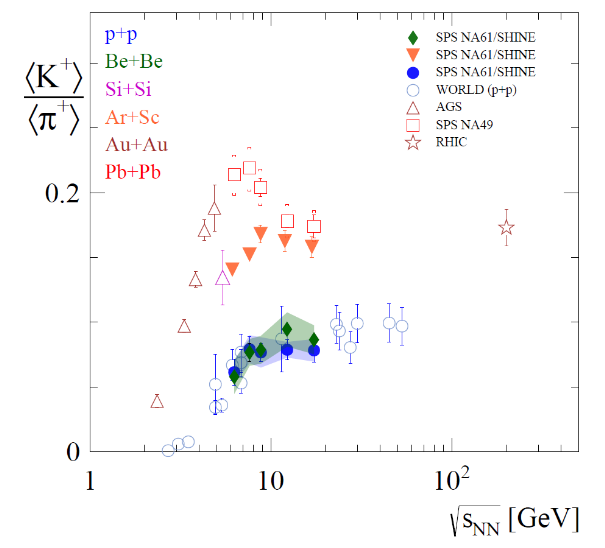}
       \put(39, 20){\scriptsize{\textbf{\textcolor{light-gray}{NA61/SHINE preliminary}}}}
    \end{overpic}   
    \caption{Ratio of yields K$^+$/$\pi^+$ in mid-rapidity (\textit{left}) and the ratio of total yields K$^+$/$\pi^+$ (\textit{right}) produced in p+p, Be+Be, Ar+Sc and Pb+Pb collisions as function of collision energy.}
    \label{fig:horn}
\end{figure}

Rapid changes of the K$^+$/$\pi^+$ yields ratio at mid-rapidity and total K$^+$/$\pi^+$ yields ratio as a function of collision energy were observed in Pb+Pb collisions by the NA49 experiment \cite{na49_onset}. These two ratios, together with new NA61/SHINE results from Be+Be and Ar+Sc collisions, are shown in Fig. \ref{fig:horn}. 
A plateau-like structure in the K$^+$/$\pi^+$ energy dependence is visible in p+p interactions. The ratio K$^+$/$\pi^+$ at mid-rapidity as well as the ratio of total yields from Be+Be collisions is close to the p+p measurements. For Ar+Sc collisions the ratio K$^+$/$\pi^+$ at mid-rapidity and the ratio of total K$^+$/$\pi^+$ yields are higher than in p+p collisions but show qualitatively similar energy dependence and no horn structure is visible.

\subsection{Proton puzzle}
\begin{figure}[h]
 \begin{overpic}[width=7.cm, trim={0 1cm 0 0.5cm }, clip]{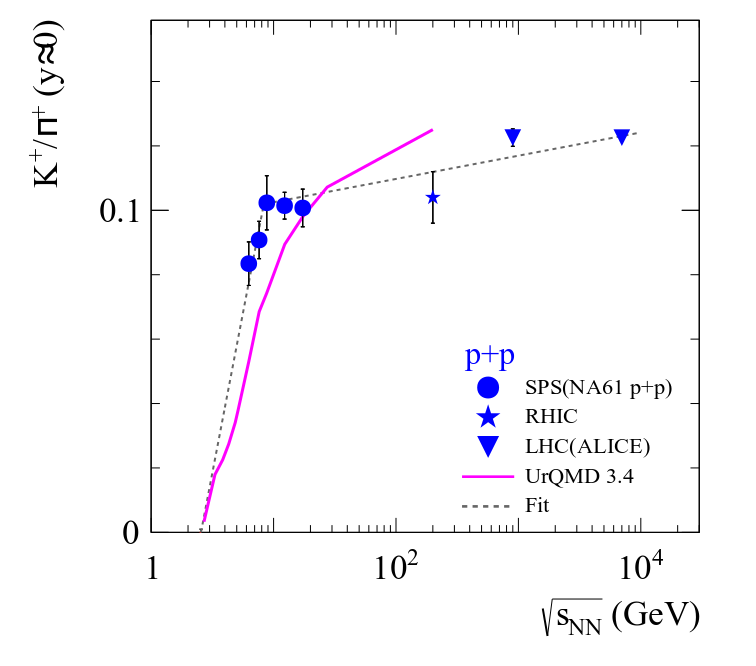}
\end{overpic} 
\begin{overpic}[width=7.cm, trim={0 1cm 0 0.5cm }, clip]{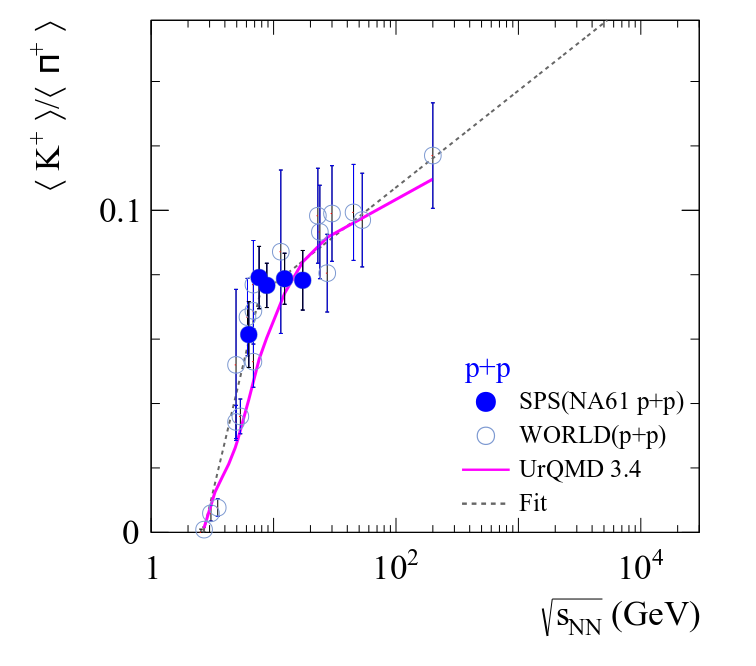}
\end{overpic}\\
    \begin{overpic}[width=7.cm, trim={0 1cm 0 0.5cm }, clip]{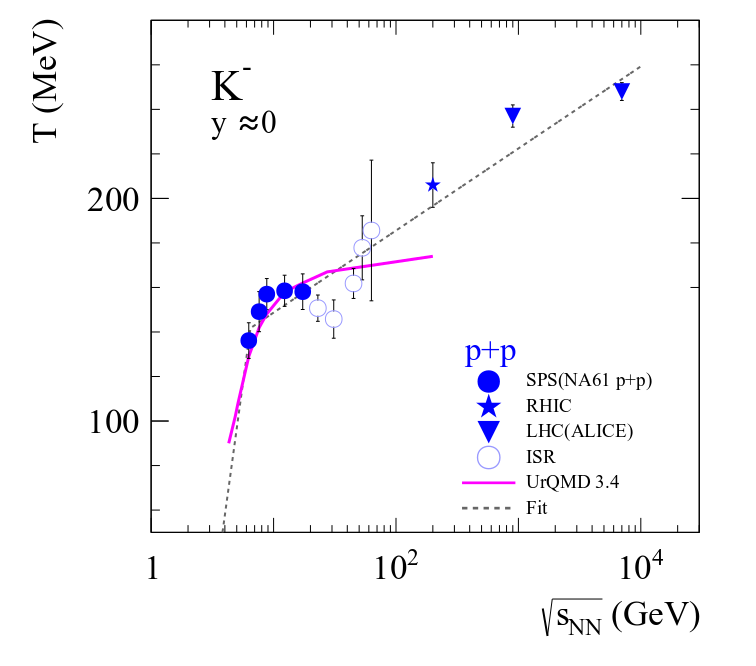}
\end{overpic} 
    \begin{overpic}[width=7.cm, trim={0 1cm 0 0.5cm }, clip]{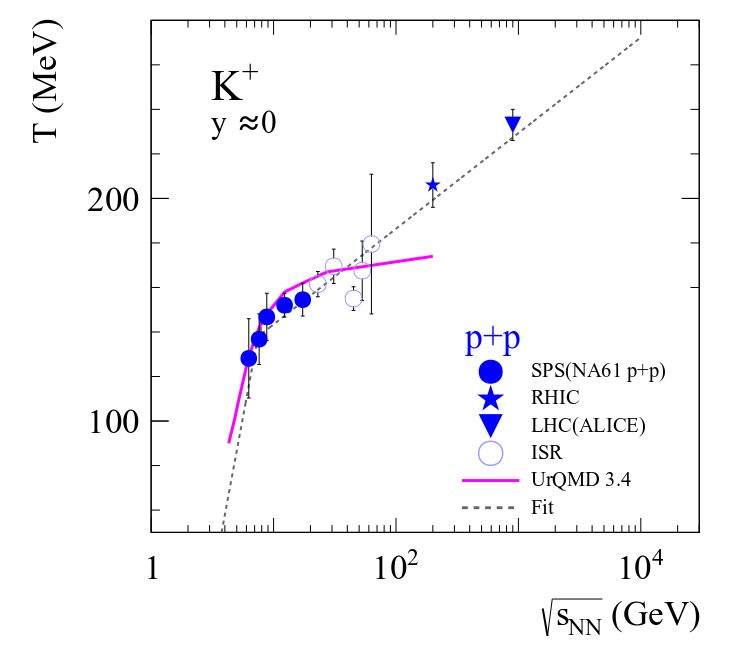}
\end{overpic} \\
\caption{Energy dependence of the K$^+$/$\pi^+$ ratio in inelastic p+p interactions in mid-rapidity (\textit{top-left}) and in the full phase-space (\textit{top-right}) as well the inverse slope parameter $T$ of transverse mass spectra in mid-rapidity for K$^-$ (\textit{bottom-left}) and K$^+$ (\textit{bottom-right}) mesons. The data are fitted by two straight lines in order to locate a position of the break in the energy dependence. The experimental results are compared with predictions of the resonance-string model, UrQMD.}
\label{fig:puzzle}
\end{figure}

Rates of increase with collision energy of the K$^+$/$\pi^+$ ratio and $T$ change sharply in  p+p interactions at SPS energies \cite{protony}. NA61/SHINE results along with available data from RHIC, LHC, ISR and others, clearly show the breaking point (see Fig.\ref{fig:horn}). To estimate the break energy between a fast rise at low energies and a plateau (or slower increase) at high energies, two straight lines were fitted to the p+p data as shown in Fig. \ref{fig:puzzle}. 
The low energy line was constrained by the threshold energy for kaon production. The fitted break energy is 8.3~$\pm$~0.6~GeV, 7.70~$\pm$~0.14~GeV, 6.5~$\pm$~0.5~GeV and 7.9~$\pm$~0.2~GeV, for the K$^+$/$\pi^+$ , $\langle$K$^+\rangle$/$\langle \pi^+\rangle$ ratios and $T$(K$^-$), $T$(K$^+$), respectively. These values are close to each other and surprisingly close to the energy of the beginning of the horn and step structures in central Pb+Pb collisions~, the transition energy being approximately 8 GeV (see Fig. \ref{fig:horn}). Figure \ref{fig:puzzle} also shows that the sharpness of the break cannot be reproduced by the resonance-string model (UrQMD) \cite{protony2}.

\subsection{Flow}
Directed flow $v_1$ was considered to be sensitive to the first-order phase transition due to strong softening of the Equation of State \cite{flow1,flow2,flow3}. The expected effect is a non-monotonic behaviour (change from positive to negative and again to positive values) of proton $dv_1/dy$ as a function of beam energy. This effect is usually referred to as the collapse of proton flow and was measured by the NA49 experiment as anti-flow of protons in peripheral Pb+Pb collisions at 40$A$ GeV/$c$ beam momentum (8.8 GeV) \cite{flow_na49}. 

\begin{figure}
\begin{minipage}{9cm}
 \includegraphics[width=9cm]{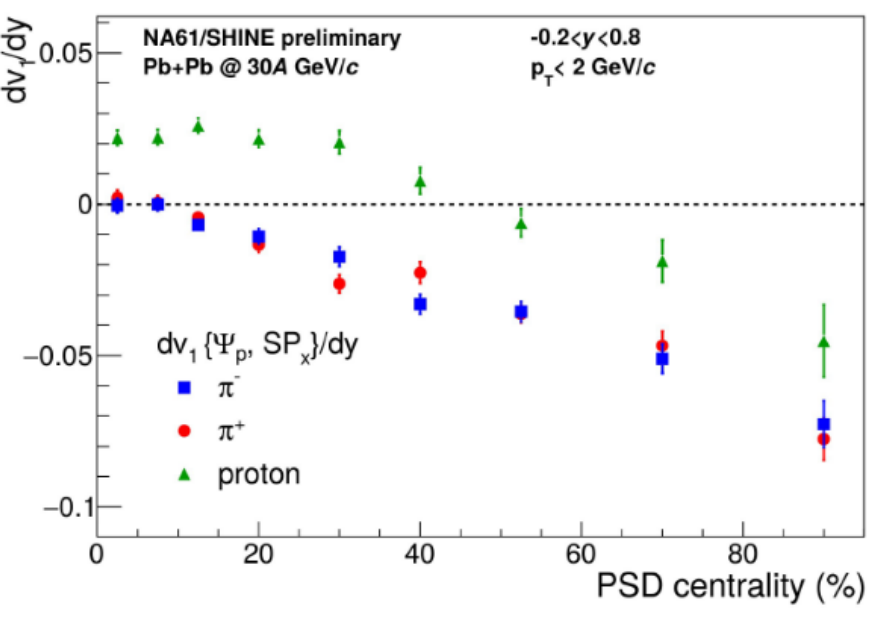}
\end{minipage}
\begin{minipage}{6.7cm}
\vspace{-0.2cm}
 \begin{overpic}[width=6.7cm]{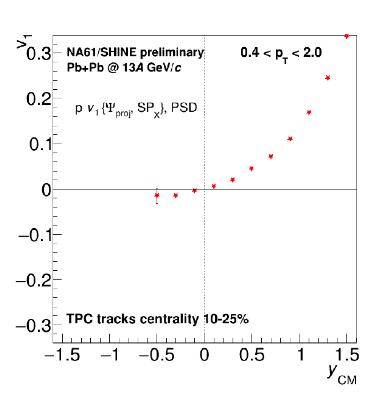}
 \end{overpic}
 \end{minipage}
 \caption{Preliminary results on centrality dependence of $dv_1 /dy$ at mid-rapidity measured in Pb+Pb collisions at 30$A$ GeV/$c$ (\textit{left}) and $v_1$ as function of rapidity measured in Pb+Pb collisions at 13$A$ GeV/$c$ (\textit{right}).}
 \label{fig:flow}
\end{figure}

In 2018 the NA61/SHINE experiment reported the first results on anisotropic flow, measured in centrality selected Pb+Pb collisions at 30$A$ GeV/$c$ beam momentum. The NA61/SHINE large-acceptance spectrometer working in the fixed target setup allows particle tracking and identification over a wide rapidity range. Moreover, the NA61/SHINE forward calorimeter, the PSD, allows to determine the collision event plane and therefore measurements of the flow coefficients relative to the spectator plane. Preliminary results on the centrality dependence of $dv_1 /dy$ at mid-rapidity, measured in Pb+Pb collisions at 30$A$ GeV/$c$, are presented in Fig. \ref{fig:flow} (\textit{left}). 
Plotted results show that the slope of pion $v_1$ is always negative, regardless of the electric charge of the particle.  In contrast, the slope of proton $v_1$ changes sign with decreasing violence of the collision, here for the centrality of the order of 50$\%$. Proton directed flow as function of rapidity for Pb+Pb collisions at 13$A$~GeV/$c$  is presented in Fig. \ref{fig:flow} (\textit{right}). Measurement at more energies are required to draw conclusions on the collapse of proton directed flow in Pb+Pb interactions~\cite{flow_na61}.

\section{Search for the critical point}
One of the main goals of the NA61/SHINE experiment is to locate the critical point (2$^\textup{nd}$ order phase transition) of strongly interacting matter. The exact location of the critical point (region) on the phase diagram is not known. It was estimated with various models and lattice calculations\cite{cp_latice1}. The latest predictions suggest the critical endpoint location in the $\mu_\textup{B} - \textup{T}$ phase diagram accessible to the studies at the SPS, and more specifically for so-called critical temperature $T_c ^{\textup{CEP}}<135-140$ MeV and critical baryon chemical potential $\mu_\textup{B}^\textup{CEP}>300$ MeV \cite{cp_latice2}.

NA61/SHINE searches the critical point using such tools like scaled factorial moments~\cite{cp_inter1}, central moments of multiplicity distributions of higher-order~\cite{cp_higher_moments} as well as intensive and strongly intensive measures~\cite{cp_quantities} of particle multiplicity and kinematic variables. Some of them will be discussed in this contribution.

\subsection{Strongly-intensive quantities}
The $2^\textup{nd}$ order of phase transition is expected to lead to enhanced fluctuations of multiplicity and transverse momentum. For their study NA61/SHINE uses strongly intensive measures, for example $\Sigma$[P$_\textup{T}$, N] \cite{cp_tobiasz}. Within the Wounded Nucleon Model (WNM), the strongly intensive quantities depend  neither on the number of wounded nucleons (W) nor on fluctuations of W. Likewise, in the Grand Canonical Ensemble they do not depend on volume and volume fluctuations. The quantity $\Sigma$ is defined as follows:
\begin{equation}
 \Sigma[P_T,N] = \frac{1}{C_{\Sigma}}\Big[ \langle N \rangle\omega_{P_T} + \langle P_T \rangle\omega_{N} - 2 \big( \langle P_TN \rangle - \langle P_T \rangle \langle N \rangle \big) \Big] ,
%
%
 \label{eq:sigma}
\end{equation}
where $\omega$ is the scaled variance of a given variable (e.g. $\omega [N] = \frac{\langle N^2 \rangle - \langle N \rangle ^2}{\langle N \rangle}  
$). The quantity is constructed in the way that it is equal to zero in the case of no fluctuations and one in the case of independent particle production.

\begin{figure}[h]
 \begin{overpic}[width=6.8cm, trim={0.3cm 0.5cm 2.8cm 1cm}, clip]{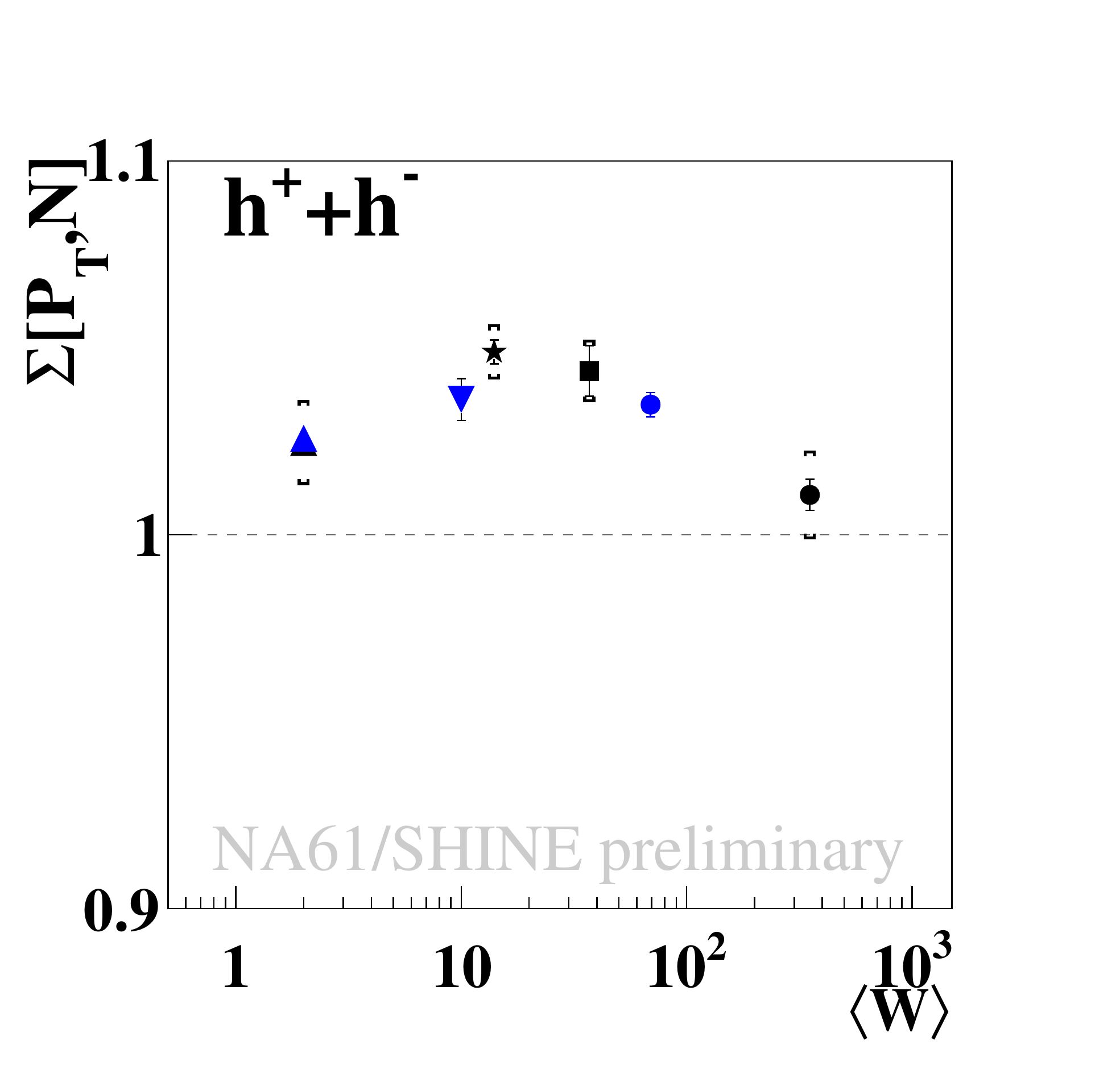}
 \put(18, 40){\scriptsize{NA49: p+p, 0-15.3\% C+C}}
 \put(18, 35){\scriptsize{0-12.2\% Si+Si, 0-5\% Pb+Pb}}
 \put(18, 30){\scriptsize{\textcolor{blue}{ NA61/SHINE: p+p, 0-5\% Be+Be}}}
 \put(18, 25){\scriptsize{\textcolor{blue}{ 0-5\% Ar+Sc}}}

\end{overpic}
\hspace{0.2cm}
 \begin{overpic}[width=9.2cm]{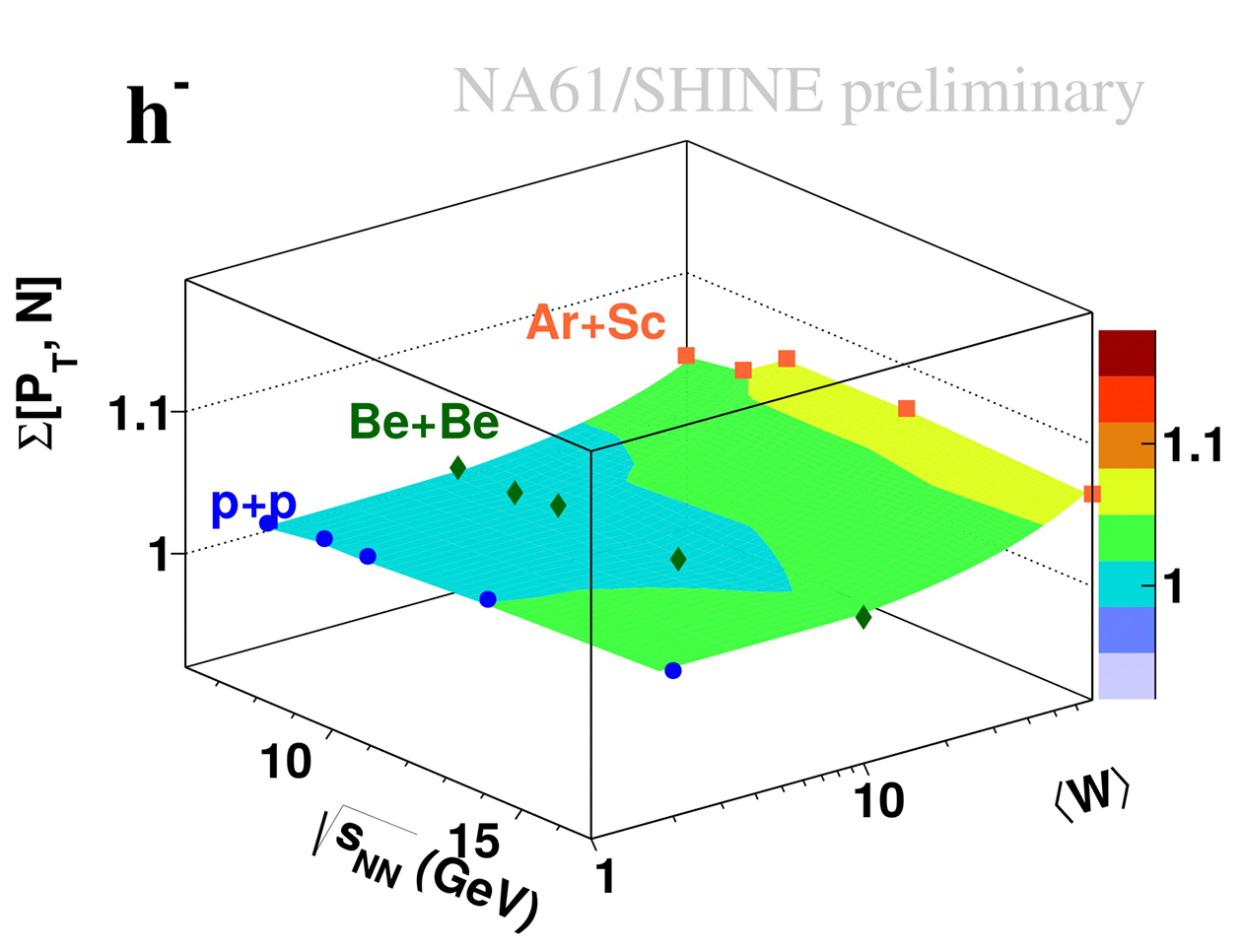}
\end{overpic}\\
\caption{$\Sigma$[P$_\textup{T}$, N] for all charged hadrons (h$^+$ + h$^−$ ) from the NA61/SHINE and NA49 \cite{sigma_na49} experiments for the NA49 acceptance ($1.1 < y_\pi <2.6$) as function of system size at 150$A$/158$A$ GeV/$c$ (\textit{left}) and $\Sigma$[P$_\textup{T}$, N] for negatively charged hadrons in inelastic p+p (blue squares) \cite{cp_tobiasz}, 0-5\% Be+Be (green diamond), and 0-5\% Ar+Sc (orange squares) collisions obtained by NA61/SHINE in the NA61/SHINE acceptance ($0.0 < y_\pi < y_{beam}$). For NA61/SHINE only statistical uncertainties are shown. }
\label{fig:sigma}
\end{figure}

The system size dependence of $\Sigma$[P$_\textup{T}$, N] at 150$A$/158$A$ GeV/$c$  from the NA61/SHINE and NA49 \cite{sigma_na49} experiments within the NA49 acceptance as function of system size (wounded nucleons) is presented in Fig. \ref{fig:sigma} (\textit{left}). 
NA49 and NA61/SHINE measurements show consistent trends in the $\Sigma$[P$_\textup{T}$, N] dependence on the system size. Finally NA61/SHINE results for $\Sigma$[P$_\textup{T}$, N] obtained in the NA61/SHINE acceptance for p+p, Be+Be and Ar+Sc collisions are
presented in Fig. \ref{fig:sigma} (\textit{right}). So far, there is no prominent structure observed which could be related to a critical point.

\subsection{Intermittency}
An intermittency signal in proton multiplicity was predicted close to the critical point. The effect is expected to manifest in local power-law fluctuations of the baryon density which can be searched for by studying the scaling behaviour of second factorial moments $F_2( M )$ with the cell size or, equivalently, with the number of cells in ($p_x$, $p_y$) space of protons at mid-rapidity \cite{cp_inter1, cp_inter3, cp_inter2}.

The transverse momentum phase-space is divided into M$\times$M equal-sized bins, and multiplicities quantify the proton distribution in individual momentum bins. The second-order factorial moment in transverse momentum space is defined as: 
\begin{equation}
 F_2(M) \equiv \frac{\left \langle \frac{1}{M^2} \sum\limits_{m=1}^{M^2} n_m (n_m-1) \right \rangle}{\left \langle \frac{1}{M^2} \sum\limits_{m=1}^{M^2} n_m \right \rangle^2},
\end{equation}
where M$^2$ means the number of bins (M bins in $p_ x$ and M bins in $p_y$ ), while $n_m$ is the number of protons in the $m$-th bin.

However, subtraction of mixed events is needed in order to remove the non-critical (trivial) background contribution of proton pairs, as following \cite{inter_old2, inter_na49}:
\begin{equation}
 \Delta F_2(M) \equiv F_2^{data}(M) - F_2^{mixed}(M) .
\end{equation}
Then the second factorial moment, $\Delta F_2(M)$ should scale according to a power-law (for M$\gg$1):
\begin{equation}
 \Delta F_2(M) \sim (M^2)^{\phi_2} .
\end{equation}

In the recent analysis of NA61/SHINE, the intermittency effects were studied in central Be+Be and centrality selected Ar+Sc collisions at 150$A$ GeV/$c$. Protons were identified with a method based on $dE/dx$ measurements and were selected with at least 90\% purity. The collision centrality was determined from the energy deposited in the PSD detector.

\begin{figure}[h]
\includegraphics[width=5cm]{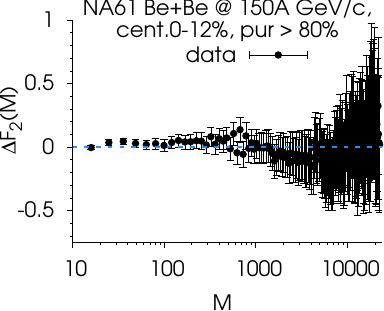}
\includegraphics[width=5cm]{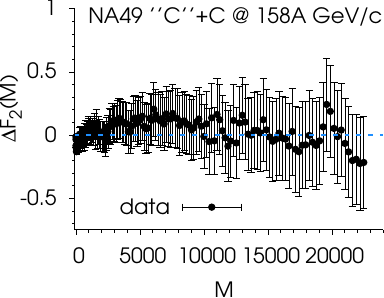}
\includegraphics[width=5cm]{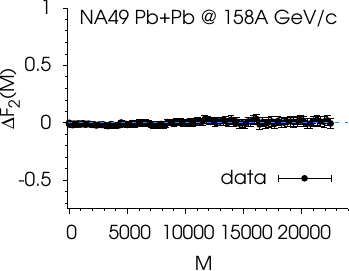}
\caption{Second factorial moment, $\Delta F_2$, for mid-rapidity protons at $\sqrt{s_\textup{NN}} \sim $17 GeV in Be+Be by NA61/SHINE (\textit{left}) \cite{inter_old2}, C+C (\textit{center}) and Pb+Pb (\textit{right}) by NA49 \cite{inter_na49}.}
\label{fig:deltaF_comparison}
\end{figure}

\begin{figure}[h]
\includegraphics[width=5cm]{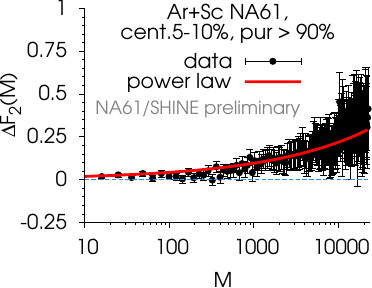}
\includegraphics[width=5cm]{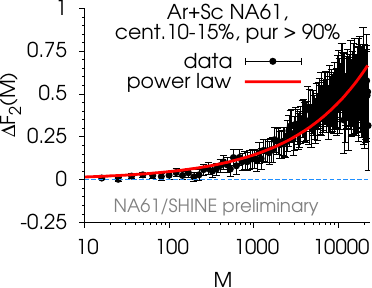}
\includegraphics[width=5cm]{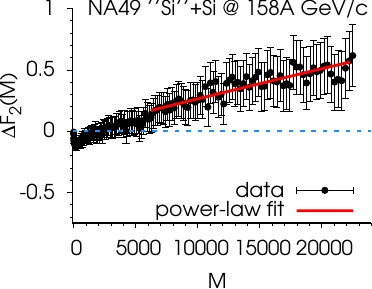}
\caption{Second factorial moment, $\Delta F_2$, for mid-rapidity protons at $\sqrt{s_\textup{NN}}$=17 GeV in Si+Si by NA49 \cite{inter_na49} (\textit{right}) and in 5-10$\%$ and 10-15$\%$ Ar+Sc by NA61/SHINE \cite{inter_old} (\textit{left} and \textit{center}).}
\label{fig:deltaF_comparison2}
\end{figure}

The NA61/SHINE results were compared with the most recently published experimental results on proton intermittency  obtained by NA49 in the same acceptance and with use of the same analysis method. Figure \ref{fig:deltaF_comparison} presents a comparison of the second factorial moment, $\Delta F_2$, for mid-rapidity protons produced at 150/158$A$ GeV/$c$ ($\sqrt{s_\textup{NN}} \sim $17 GeV) in Be+Be by NA61/SHINE \cite{inter_old2}, C+C and Pb+Pb by NA49 \cite{inter_na49}.
All results mentioned above do not indicate the power law behaviour, in contrast to the results presented in fig. \ref{fig:deltaF_comparison2}. They show the same quantity in centrality selected: 5-10$\%$ and 10-15$\%$ Ar+Sc collisions by NA61/SHINE \cite{inter_old2}, in which a hint of power law behaviour can be observed, and in Si+Si collisions by NA49 \cite{inter_na49}, in which the intermittency effect in proton multiplicity is clearly visible.

\begin{figure}[h]
\includegraphics[width=7.7cm]{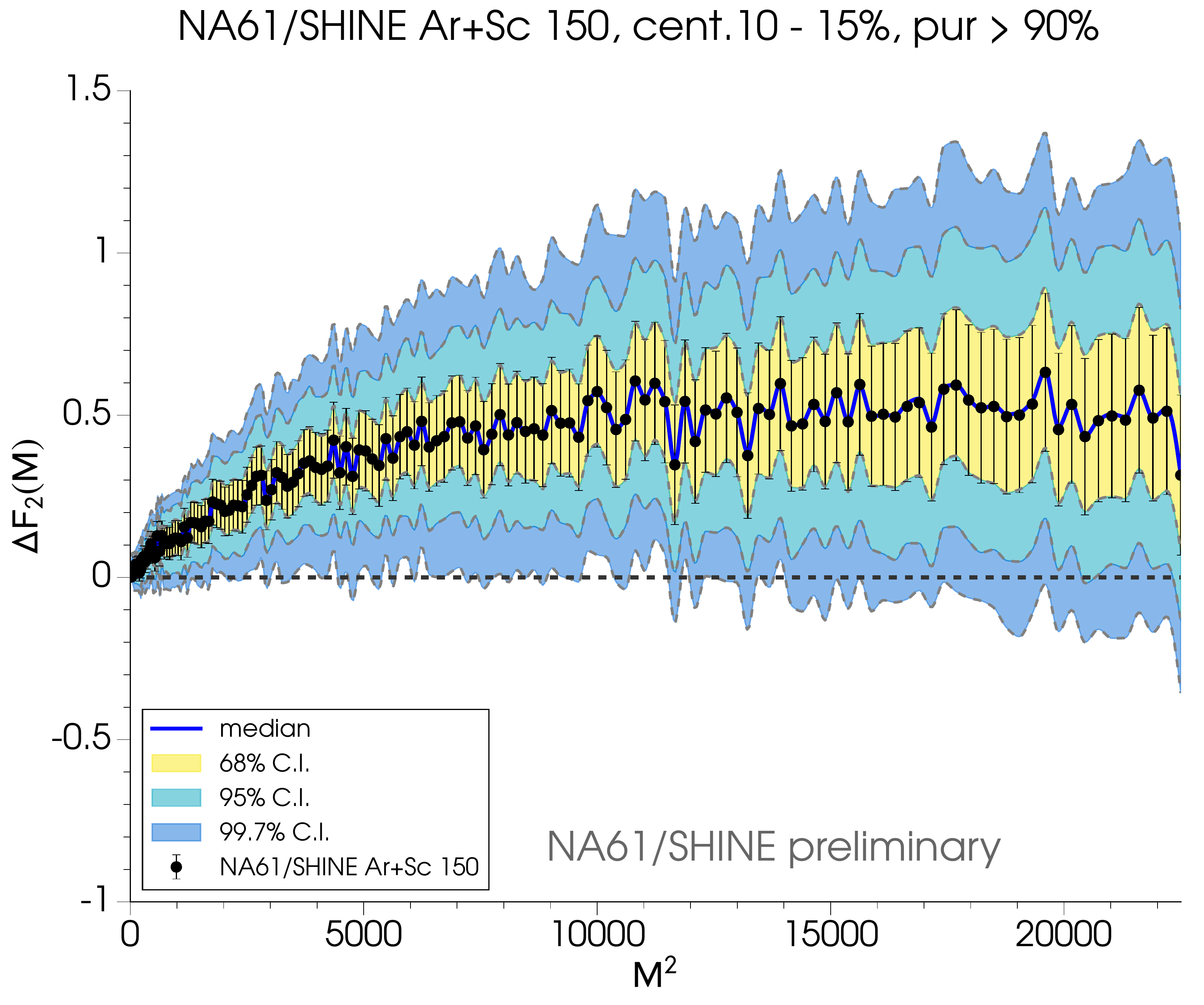}
\includegraphics[width=7.7cm]{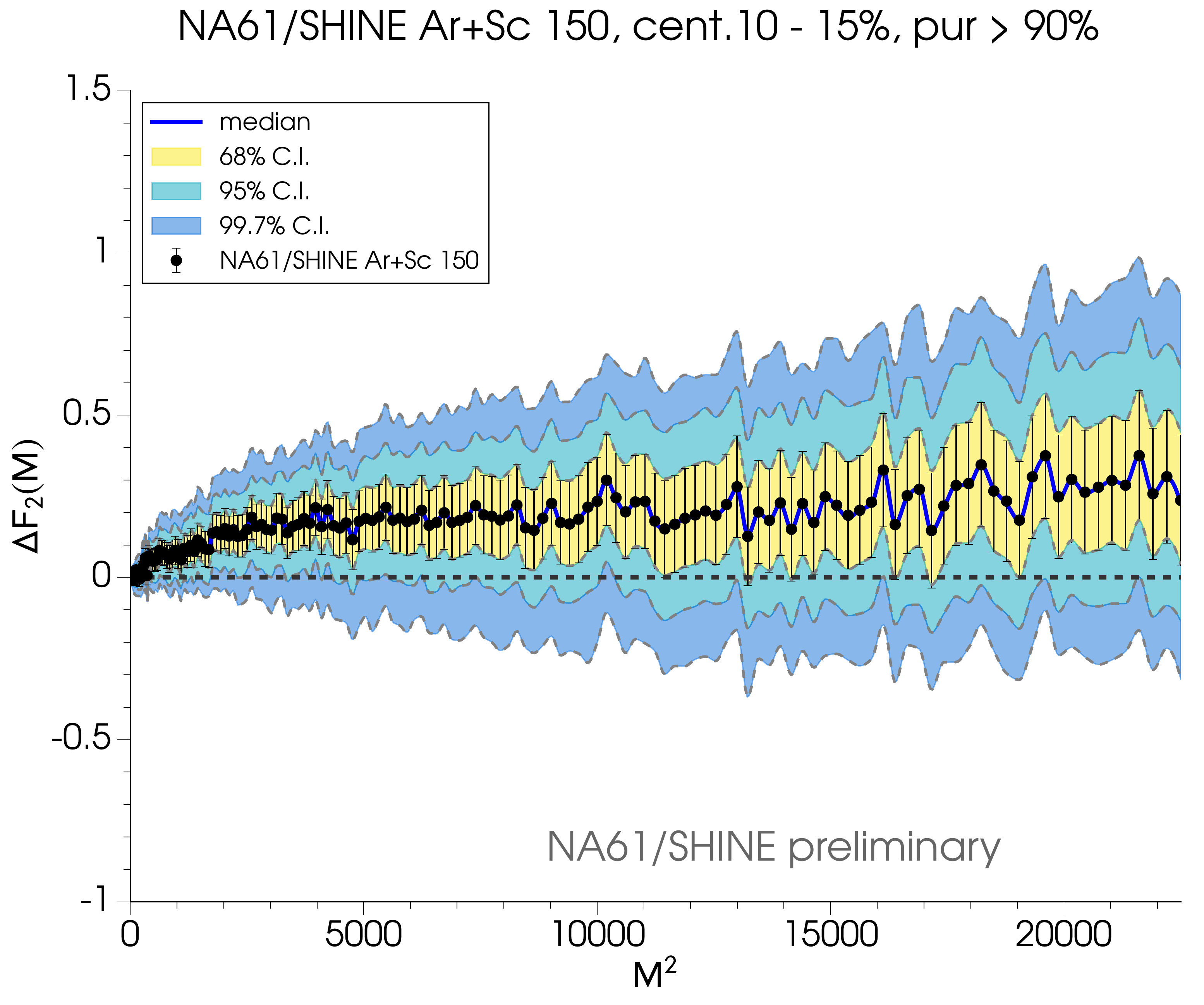}
\caption{Preliminary results on $\Delta F_2(M)$ of mid-rapidity protons measured in 10-15\% central Ar+Sc collisions at 150$A$ GeV/$c$ with lower (\textit{left}) and higher (\textit{right}) statistics \cite{inter_new}.}
\label{fig:inter}
\end{figure}

The results on proton intermittency in Ar+Sc collisions at 150$A$ GeV/$c$ have been revised for the higher statistics of events and presented again in fig. \ref{fig:inter}.
Left hand side shows results released in 2018 \cite{inter_old2, inter_old} and presented at the ICNFP 2019. These results indicate an increase of $\Delta F_2(M)$ values with number of bins M, which may be connected with a proximity to the CP. Right hand side shows the same results but with higher statistics (208k events vs 143k events) where $\Delta F_2(M)$ signal is weaker and show no strong evidence for the critical point observation \cite{inter_new}.

\section{Summary and outlook}
This contribution focuses on recent results from the NA61/SHINE strong interactions program aiming to study the onset of deconfinement and search for the critical point of strongly interacting matter. 

Results on charged kaon spectra in Ar+Sc collisions at 19$A$--150$A$ GeV/$c$ were presented. In particular, the inverse slope parameter of the transverse mass distribution and the charged kaon to pion multiplicity ratio were discussed. 
The results on K$^+$/$\pi^+$ ratio in mid-rapidity and the ratio of total K$^+$/$\pi^+$ yields obtained for Ar+Sc collisions are in-between results from p+p/Be+Be and Pb+Pb collisions, but qualitatively show similar energy dependence to small systems and does not exhibit the signature, which could be attributed to the onset of deconfinement (horn).

Results on energy dependence of K$^+$/$\pi^+$ ratio and the inverse slope parameter of kaon $p_\textup{T}$ spectra in inelastic p+p interactions were presented together with available data from RHIC, LHC, ISR and others. The rates of these observables are rapidly changing with the collision energy and clearly show the breaking point:  8.3~$\pm$~0.6~GeV, 7.70~$\pm$~0.14~GeV, 6.5~$\pm$~0.5~GeV and 7.9~$\pm$~0.2~GeV, for the K$^+$/$\pi^+$, $\langle$K$^+\rangle$/$\langle \pi^+\rangle$ ratios and $T$(K$^-$), $T$(K$^+$), respectively. These values are surprisingly close to the energy of the beginning of the horn and step structures in central Pb+Pb collisions-the transition energy being approximately 8 GeV. The sharpness of the break cannot be reproduced by the resonance-string model (UrQMD) and might be a hint of critical effects (e.g. onset of deconfinement) in inelastic p+p interactions.

Results on directed flow of protons in semi-peripheral Pb+Pb collisions at 13$A$ GeV/$c$ and 30$A$ GeV/$c$ were presented. Still, results in broader energy range, 40$A$ and 150$A$ GeV/$c$, are expected and needed to draw conclusions on the collapse of proton directed flow in Pb+Pb collisions. So far, no prominent structures, which can be attributed to the onset of deconfinement, were observed by NA61/SHINE.

Current NA61/SHINE results on the search for CP via strongly intensive quantities, obtained by varying number of wounded nucleons and collision energy, reveal no prominent structure, which could be related to a critical point. Nevertheless, the NA61/SHINE measurements show consistent trends in the $\Sigma$[P$_\textup{T}$, N] dependence on the system size to the one obtained by NA49 experiment.

At the present moment, studies of midrapidity proton intermittency in central and semi-central Ar+Sc collision, with enlarged statistics compared to previously reported results, show no statistically significant signal for the CP. NA61/SHINE continues the analysis of data recorded within its scan in collision energy and nuclear mass number of colliding nuclei. New analysis methods are being developed. Thus many more results on the search for the critical point of strongly interacting matter are expected in the years to come.

Currently, NA61/SHINE undergoes a significant upgrade to prepare for the extension of the scientific programme to open charm production measurements in Pb+Pb collisions in the SPS energy range.

\end{document}